\documentclass[twocolumn,floats,prl,showpacs]{revtex4-1}

\usepackage{amsmath}
\usepackage{amsfonts}

\usepackage{tikz}

\newcommand {\be}{\begin{equation}}
\newcommand {\ee} {\end{equation}}
\newcommand {\bea}{\begin{eqnarray}}
\newcommand {\eea} {\end{eqnarray}}
\newcommand{\non}{\nonumber}

\begin{document}


\title{Proof of single-replica equivalence in short-range spin glasses}

\author{C.M.~Newman$^{1,2}$, N. Read$^{3,4}$, and D.L. Stein$^{1,2,5,6,7}$}
\affiliation{$^1$Courant Institute of Mathematical Sciences, New York University, New York, NY 10012, 
USA\\
$^2$NYU-ECNU Institute of Mathematical Sciences at NYU Shanghai, 3663 Zhongshan Road North, Shanghai, 
200062, China\\
$^3$Department of Physics, Yale University, P.O. Box 208120, New Haven, Connecticut 06520-8120, USA\\
$^4$Department of Applied Physics, Yale University, P.O. Box 208284, New Haven, Connecticut 06520-8284, 
USA\\
$^5$Department of Physics, New York University, New York, NY 10012, USA\\
$^6$NYU-ECNU Institute of Physics at NYU Shanghai, 3663 Zhongshan Road North, Shanghai, 200062, 
China\\
$^7$Santa Fe Institute, 1399 Hyde Park Rd., Santa Fe, NM 87501, USA}

\date{January 25, 2023}

\begin{abstract}
We consider short-range Ising spin glasses in equilibrium at infinite system size, and prove that,
for fixed bond realization and a given Gibbs state drawn from a suitable metastate, each translation- and locally-invariant function 
(for example, self-overlaps) of a single pure state in the decomposition of the Gibbs state takes the same value for all the 
pure states in that Gibbs state. We describe several significant applications to spin glasses.
\end{abstract}


\maketitle

The nature of the spin-glass (SG) phase in classical finite-dimensional short-range models remains one of the outstanding
unsolved problems in statistical mechanics.  Although important fundamental questions 
remain open, considerable analytical and numerical progress has been made, especially on the rigorous theory of mean-field 
SG's~\cite{AC98,GG98,GT02,Guerra03,Talagrand03a,Talagrand03b,Talagrand05,panch} and short-range SG's
(for a recent review, see Ref.~\cite{NRS22}). For the mean-field case, which corresponds to infinite-range models 
such as the Sherrington-Kirkpatrick (SK) Ising Hamiltonian~\cite{SK75}, most of the fundamental problems
have been solved by the replica symmetry breaking (RSB) theory~\cite{Parisi79,Parisi83,MPSTV84a,MPSTV84b,MPV87}. 

For the short-range case, in which we will focus on Ising spins ($s_{\bf x}=\pm 1$ for all sites $\bf x$) and infinite system 
size, there is an unresolved controversy about whether the low-temperature phase involves many ordered or ``pure'' states
as in RSB, or only one or two, as in the  scaling-droplet (SD) picture~\cite{Mac84,BM85,FH88b,FH86}. Rigorous 
results have been obtained using {\em metastates}~\cite{AW90,NS96c,NS97,NSBerlin}; a metastate is a probability distribution 
on equilibrium (i.e.\ Gibbs) states, with covariance properties we describe below. In the SD picture, the metastate is trivial 
(i.e.\ supported on a single Gibbs state), while for RSB behavior the metastate is 
necessarily nontrivial~\cite{NS96c,NS97,NSBerlin,NS02,NS03b,Read14,NRS22}, and a Gibbs state in its support is a nontrivial 
mixture of many pure states (if there is global spin-flip symmetry under $s_{\bf x}\to-s_{\bf x}$ 
for all ${\bf x}$, these are not all related by symmetry).

In this paper, we  establish a further necessary property of pictures with Gibbs states that are nontrivial mixtures of pure states. 
Loosely, for systems {\em without} spin-flip invariance, there is no macroscopic order parameter that can distinguish 
between the pure states in a Gibbs state; i.e., for given bonds and Gibbs state (drawn from a metastate,
which can be trivial or nontrivial), all pure states in the Gibbs state ``look alike'', in that
each macroscopic property (defined precisely later) defined for any single pure state takes the same 
value in all the pure states.  For example, all pure states in a given Gibbs state have the same self-overlap, 
magnetization, and internal energy density. Similarly, {\em with} spin-flip invariance, 
pure states cannot be distinguished from one another by {\it flip-invariant\/} order parameters (note that magnetization 
is not flip-invariant).  We call this property ``single-replica equivalence''. 
(A similar statement that self-overlaps almost surely take a single value in {\em infinite}-range models, assuming the 
Ghirlanda-Guerra identities~\cite{GG98} hold, was proved in Ref.~\cite{Panchenko10a}.)  
This result has a number of immediate applications that we describe later.
For technical reasons, the proof of our result is for 
models with interactions within groups of $p$ spins, for all $p$ (or all even $p$); the case of only nearest-neighbor pair 
interactions is not included, but can be approached arbitrarily closely.

Single-replica equivalence is so-named because of its similarity to replica 
equivalence~\cite{Parisi04}. Here the term ``replica'' refers to real replicas, i.e., pure states drawn from some 
distribution. Replica equivalence asserts that functions of overlaps of distinct replicas are 
independent of the choice of one of the replicas; this is not the property that we discuss,
but may possibly be related. 

We now define notations and review concepts that will be needed in what follows.  
The sites $\bf x$ lie in the $d$-dimensional cubic lattice $\mathbb{Z}^d$, and we define 
$s=(s_{\bf x})_{{\bf x}\in\mathbb{Z}^d}$. Let $X$ denote a nonempty finite set of distinct sites, 
$\cal X$ the set of all such $X$, and $s_X=\prod_{{\bf x}\in X}s_{\bf x}$. A general Hamiltonian is then
\begin{equation}
\label{eq:general}
H_J(s)=-\sum_{X\in{\cal X}} J_X s_X\, ,
\end{equation}
where $J=(J_X)_{X\in{\cal X}}$ is an indexed set of independent random variables (bonds), one associated with each 
$X\in{\cal X}$, so the joint distribution $\nu(J)$ of $J_X$ for all $X$ is a translation-invariant product (over $X$) 
distribution; we write expectation under $\nu$ as ${\bf E}\cdots$. We define a ``mixed $p$-spin model'' of this form 
(``mixed'' means the sum is over all $X\in{\cal X}$, and $p$ denotes values of $|X|$)
to be (I) ``short range'' if  $\sum_{X:{\bf x}\in X}{\bf E}|J_X|<\infty$ for any $\bf x$ (see 
e.g.~Ref.\ \cite{Read22}), and (II) ``n.i.p.''\ if, for every $X$ such that ${\rm Var}\, J_X>0$ (possibly infinite), 
there are {\em no isolated points} in the support of the marginal distribution for $J_X$ [e.g.\ the marginal is continuous 
(i.e.\ atomless)]. 
For spin-flip invariance of $H_J$ under $s\to-s\equiv (-s_{\bf x})_{{\bf x}\in\mathbb{Z}^d}$ we impose also (III) $J_X=0$ 
for $|X|$ odd. The familiar EA~Hamiltonian~\cite{EA75} is a special case of these models in which $J_X=0$ if $X$ is neither 
a nearest-neighbor pair nor a single site. For a SG one typically assumes $J_X$ has mean zero (except possibly for $|X|=1$, 
the single-site magnetic field terms), but the mean-zero assumption is neither required nor assumed in the theorems 
and proofs below. 

States $\Gamma$ [i.e., probability distributions $\Gamma(s)$ on configurations $s$]  are uniquely determined
by the values of the expectations $\langle s_X\rangle_\Gamma$ in $\Gamma$ as $X$ runs through $\cal X$. 
An (infinite-volume) Gibbs state
is defined for a given short-range Hamiltonian, such as $H_J(s)$, and for fixed temperature $T$ ($0<T<\infty$) 
as a state that obeys the Dobrushin-Lanford-Ruelle conditions \cite{Georgii88,Simon92}. A convex combination (i.e.\ a mixture)
of Gibbs states is again a Gibbs state.

A Gibbs state may be either pure or mixed. A pure state is a Gibbs state that is extremal, i.e.\ not expressible  
as a mixture of other Gibbs 
states.  Equivalently,  it obeys a strong clustering property~\cite{Georgii88,Simon92} that implies decay of connected 
correlations to zero. Distinct pure states put all their probability on disjoint sets of spin configurations 
\cite{Georgii88,Simon92}. We will denote pure states by $\Gamma_\alpha$ and expectation in $\Gamma_\alpha$ as 
$\langle\cdots\rangle_\alpha$ ($\alpha$ is an index). Any Gibbs state~$\Gamma$ can be expressed, 
or ``decomposed'', as a unique mixture of pure states \cite{Georgii88}, that is
\be
\label{eq:decomp}
\Gamma =\sum_\alpha w_\alpha \Gamma_\alpha
\ee
for a set of non-negative weights $w_\alpha=w_\Gamma(\alpha)$ that 
obey $\sum_\alpha w_\alpha=1$ (i.e.\ probabilities) and which depend on $J$ and 
$\Gamma$. Eq.\ (\ref{eq:decomp}) corresponds to a countable decomposition, but our results hold in 
the general case, where every sum $\sum_\alpha w_\alpha\cdots$ with weights $w_\alpha$ becomes an integral 
$\int dw_\Gamma(\alpha)
\cdots$ with probability measure $dw_\Gamma(\alpha)$. A spin-flip transformation sends any state $\Gamma$ to a state 
$\overline{\Gamma}$, defined by $\overline{\Gamma}(s)=\Gamma(-s)$.
$\overline{\Gamma}=\Gamma$ if and only if $\langle s_X\rangle_\Gamma=0$ whenever $|X|$ is odd. 
Spin-flip symmetry of $H_J$ implies that for each pure state $\Gamma_\alpha$ there is a flipped pure state
$\Gamma_{\overline{\alpha}}=\overline{\Gamma_\alpha}$, and that for a flip--invariant Gibbs state we have 
$w_{\overline{\alpha}}=w_\alpha$ for all $\alpha$.

Two other types of transformations will be important. The first type are translations:
if all bonds in a given $J$ are translated by a fixed amount, then the same translation applied to 
any Gibbs state $\Gamma$ for $J$ produces a corresponding Gibbs state for the translated $J$. The second type are 
local transformations: for any $\Delta J_X\neq0$ for finitely many $X$, a state $\Gamma$ transforms to a state 
$\Gamma'$ defined by~\cite{AW90,NS98}
\be
\langle\cdots\rangle_{\Gamma}\to\langle\cdots\rangle_{\Gamma'}=\frac{\langle\cdots 
e^{\beta\sum_X\Delta J_Xs_X}\rangle_\Gamma}{\langle e^{\beta\sum_{X'}\Delta J_{X'}s_{X'}}\rangle_\Gamma},
\label{stateloctran}
\ee
where $\beta=1/T$. When $\Gamma$ is a {\em pure} state $\Gamma_\alpha$ for $H_J$, the locally transformed state is a 
pure state $\Gamma_\alpha'$ for $H_{J+\Delta J}$, 
so we can use  the same labels $\alpha$. 

More generally, states $\Gamma$ and $\Gamma'$ are 
related as in~(\ref{stateloctran}) and are Gibbs states for $H_J$ and $H_{J+\Delta J}$, respectively, if and only if 
they are mixtures of pure states $\Gamma_\alpha$, $\Gamma_\alpha'$ for the respective
Hamiltonians with respective weights $w_\alpha$, $w_\alpha'$ related by~\cite{AW90,NS98}
\be
w_\alpha'=\frac{r_\alpha w_\alpha}{\sum_\gamma r_\gamma w_\gamma},
\label{wloctran}
\ee
where 
\be
r_\alpha =\langle e^{\beta\sum_X\Delta J_X s_X}\rangle_\alpha\,.
\ee

Our main objects of interest are {\it invariant observable properties\/} of pure states. We define an invariant 
observable~$O(J,\Gamma_\alpha)$ to be a (Borel measurable) function of $(J,\Gamma_\alpha)$ that is invariant under 
both translations and local transformations. When spin-flip symmetry is present, we consider local changes in $J_X$ only for 
$|X|$ even, and we can consider observables that are also invariant under a spin-flip transformation of $\Gamma_\alpha$.

Examples include translation averages of spin expectations. Define the translation of a site $\bf x$ by a vector ${\bf x}'$ to be 
$\tau_{{\bf x}'}{\bf x}={\bf x}+{\bf x}'$; for a set 
$X=\{{\bf x}_i:i=1,\ldots,p\}$ of sites, $\tau_{{\bf x}'}X$ is defined in the obvious way.  Also, for $W$ a positive 
odd integer, define $\Lambda_W\subset \mathbb{Z}^d$ to be a hypercube of side $W-1$, centered on the origin, so 
that $|\Lambda_W|=W^d$ sites. For a function $f_X$, its 
translation average ${\rm Av}\,f_X$ is
\be
{\rm Av}\,f_X = \lim_{W\to\infty}\frac{1}{W^d}\sum_{{\bf x}'\in \Lambda_W}f_{\tau_{{\bf x}'}X},
\ee
provided the limit exists. 

Postponing the latter issue for a moment, examples of invariant observables for a $\Gamma_\alpha$ are: 
\newline
(i) the magnetization per site, ${\rm Av}\,\langle s_{\bf x}\rangle_\alpha$ (here $X=\{{\bf x}\}$ and 
$f_X=\langle s_{\bf x}\rangle_\alpha$), and generalizations to all $s_X$ in place of a single spin $s_{\bf x}$; 
\newline
(ii) the EA single-site quadratic self-overlap 
${\rm Av}\,\langle s_{\bf x}\rangle_\alpha^2$, the two-site or edge self-overlaps ${\rm Av}\,\langle s_{\bf x} s_{\bf y}
\rangle_\alpha^2$ (for which $X=\{{\bf x},{\bf y}\}$), and their generalizations to all $s_X$; 
\newline
(iii) more general forms, involving the overlaps of all degrees [(i) and (ii) are special cases],
\be
{\rm Av}\,\prod_{i=1}^n\langle s_{X_i}\rangle_\alpha, 
\label{genover}
\ee
where $X_i$, $i=1$, \ldots, $n$ are finite sets, $X=\cup_i X_i$, and the translation average is over 
simultaneous translations of all $X_i$. 
\newline
Spin-flip invariant examples include all those in (ii), and those 
in~(\ref{genover}) if $\sum_i|X_i|$ is even.

Other examples are: (iv) parts of the internal energy density, with $f_X=-J_X \langle 
s_X\rangle_\alpha$ for each $X$, and the internal energy density itself; 
(v) the free energy density, and hence using (iv) the entropy density also; all of (iv), (v) are spin-flip invariant 
whenever $H_J$ is. 

To make further progress we introduce metastates. 
A metastate $\kappa_J(\Gamma)$ is a probability distribution on states $\Gamma$ for given $J$, such that a state 
drawn from it is a Gibbs state for $J$, with $\nu\kappa_J$-probability one [i.e.\  $\nu\kappa_J$-almost 
every $(J,\Gamma)$]; write ${\bf E}_{\kappa_J}\cdots$ for expectation under $\kappa_J$. 
Metastates were originally constructed to describe asymptotically-large finite-size systems in 
equilibrium~\cite{AW90,NS96c,NS97,NSBerlin}. They are particularly useful for systems with chaotic size 
dependence~\cite{NS92}, which may prevent directly taking the thermodynamic limit with 
bond-independent boundary conditions (BCs). Metastates using periodic BCs in the finite-size 
systems are covariant under both translations and local transformations~\cite{AW90,NS96c,NS97,NSBerlin}.
Covariance states that, if $\theta$ denotes either a translation or a local change of~$J$, and also the corresponding 
transformation of a state $\Gamma$, then $\kappa_{\theta J}(\Gamma)=\kappa_J(\theta^{-1}\Gamma)$. That is, under 
a transformation of $J$ of either type, the weight in $\kappa_J$ flows to corresponding transformed Gibbs states. 
These properties are crucial in what follows.

For $H_J$ with spin-flip symmetry, we require a metastate such that any Gibbs state drawn from 
it is spin-flip invariant. This is automatic when a spin-flip invariant BC (e.g.\ periodic) is used in the construction.

To show invariance for an observable in (i)--(iv) above, we use translation invariance of $\nu$, 
translation covariance of $\kappa_J$, and also translation covariance of $w_{\Gamma}(\alpha)$ under translations of 
$(J,\Gamma)$, which follows from the translation property of $\Gamma$. Together these imply that the probability 
distribution $\nu(J)\kappa_J(\Gamma) w_\Gamma(\alpha)$ on $J$, $\Gamma$, and $\Gamma_\alpha$ is translation 
invariant. For any function $f_X$ of $J$, $\Gamma_\alpha$ for given $(J,\Gamma)$ such that 
${\bf E}\,{\bf E}_{\kappa_J}\!\int\! dw_\Gamma(\alpha)|f_X|<\infty$, it follows directly from the ergodic theorem for 
translations \cite{Georgii88} that ${\rm Av}\,f_X$ exists and is translation invariant, for $\nu\kappa_J w_\Gamma$-almost every
$(J,\Gamma,\Gamma_\alpha)$. Invariance under local 
transformations then also holds, because the translation average involves a sum over ${\bf x}'\in\Lambda_W$, 
and by the clustering property of pure states the change in each thermal average is arbitrarily small except for 
a fraction of ${\bf x}'$s that tends to zero as $W\to\infty$. We discuss the free energy density after Theorem 1.

We can now formulate a full statement of our result:

{\bf Theorem 1}: Consider a short-range n.i.p.\ mixed $p$-spin model with ${\rm Var}\,J_X>0$ for all $X\in{\cal X}$, 
and an invariant observable property $O(J,\Gamma_\alpha)$, where the pure state $\Gamma_\alpha$ appears in the
decomposition~(\ref{eq:decomp}) of a Gibbs state $\Gamma$ drawn from a metastate $\kappa_J$, with $J$ drawn from $\nu$.
Then for $\nu$-almost every $J$ and $\kappa_J$-almost every $\Gamma$, $O(J,\Gamma_\alpha)=O(J,\Gamma_{\alpha'})$ 
for $w_\Gamma\times w_\Gamma$-almost every pair of pure states $\alpha$, $\alpha'$ in the decomposition of $\Gamma$. 
In the spin-flip-invariant case, the conditions are the same except that ${\rm Var}\,J_X>0$ for all $X\in{\cal X}$ with $|X|$ even; 
then the statement holds for invariant observables $O$ that are spin-flip invariant.

{\bf Remarks}: a) We emphasize 
that the result does not say that the invariant observable takes 
the same value in pure states in the decomposition of {\em different\/} Gibbs states.
(If for two Gibbs states there is a set of pure states having nonzero
weight in both, then all the pure states in both decompositions must have the same value of the 
observable.) 

b) The conventional picture of a first-order phase transition (FOT) is that at an FOT point two or more
pure states occur and differ in the values of some $O$s. By Theorem 1, if two or more such pure states occurred 
(for flip-invariant $O$, if there is spin-flip symmetry), each with nonzero $\nu\kappa_J w_\Gamma$ probability, 
then $\kappa_J$ would be nontrivial ($\nu$-almost surely), with the different values of $O$ segregated in distinct Gibbs 
states in the support of $\kappa_J$. The way this arises in the cases of $O$s as in examples (i), or the energy or entropy 
densities, is that at the FOT point, for each finite size one or other of the two states is favored by sample-to-sample 
fluctuations of disorder that couple to $O$ locally, so in the limit the $\kappa_J$-probability of
a mixture of the two is zero. One example is the random-field Ising ferromagnet, 
in which there is no spin-flip symmetry, and for $d>2$ at low $T$ there are two pure states with opposite magnetization 
\cite{IM75}, while $\kappa_J$ is nontrivial \cite{AW90,chatterjee_15}; others are FOTs with nonzero latent heat \cite{hb}, 
in which the local transition temperature fluctuates \cite{harris}, though in those it is less clear whether both the high and low 
temperature states are present in $\kappa_J$ at the FOT.

c) We define the free energy density of a Gibbs state $\Gamma$ as 
$\lim_{W\to\infty}F_W/W^d$ (if it exists), where 
$e^{\beta F_W+W^d\ln 2} =\langle e^{\beta H_W}\rangle_\Gamma$ and $H_W =-\sum_{X:X\cap\Lambda_W
\neq\emptyset}J_Xs_X$ \cite{AW90}. The existence of the limit for short-range $H_J$ can be proved in a similar way 
as that of the usual thermodynamic limit \cite{vuillermot,ledrappier,pasfig,vHP,vEvH_83,Zegarlinski91}; 
the invariance properties then follow easily. The proof shows directly that 
the free energy density is $\nu$-almost surely a constant, independent of both $J$ and $\Gamma$, for any Gibbs 
(not just any pure) state $\Gamma$, so for this observable our result is not needed. This approach extends to the 
magnetization and entropy densities by taking derivatives after the $W\to\infty$ limit, but the derivatives
may be undefined at an FOT point, unlike in our approach above.

d) Theorem 1, together with the $L^1$ ergodic theorem \cite{breiman_book}, further implies such results as that,
for each $X$,
\bea
\lefteqn{\lim_{W\to\infty}{\bf E}\,{\bf E}_{\kappa_J}\int\! dw_\Gamma(\alpha)\left|\frac{1}{W^d}\sum_{{\bf x}'
\in \Lambda_W}\big(J_{\tau_{{\bf x}'}X} \langle s_{\tau_{{\bf x}'}X}\rangle_\alpha\right.}\qquad\qquad\;\;\;&&\non\\
&&\qquad\qquad\qquad\;\;\left.\vphantom{\frac{1}{W^d}\sum_{{\bf x}'\in \Lambda_W}}
{}- J_{\tau_{{\bf x}'}X} \langle s_{\tau_{{\bf x}'}X}\rangle_\Gamma\big)\right|=0\qquad
\eea
{\em even} at an FOT point; these yield stronger statements of important identities 
\cite{AC98,GG98} for short-range SGs.

e) Equality of self-overlaps in the pure states in a Gibbs state is frequently used as a hypothesis, for example, 
in Refs.~\cite{NS09,ANS15}, and that is now justified by Theorem 1.
An extension of the result, under appropriate conditions, that gave equality of self-overlaps
for {\em all} pure states in all Gibbs states in the metastate would agree with RSB~\cite{NRS22}.

f) For technical reasons, the proof of Theorem 1 assumes $\nu$ is n.i.p.\ with ${\rm Var}\,J_X>0$ for all $X\in{\cal X}$ 
(or all even $X$), which excludes the EA model. Note, however, that 
${\rm Var}\, J_X$, while required to be nonzero for all $X$ (or all even $X$), could be 
taken to be arbitrarily small for all but nearest-neighbor pairs (in this example), and rapidly decaying in $X$. 
One would expect the effect of adding a 
very small perturbation (not changing the symmetry) to the EA model to have little physical effect; thus the result may 
hold more generally. Alternatively, one can 
argue that there was no physical reason to assume only nearest-neighbor interactions, as multispin interactions certainly 
occur generically in nature, even if they are usually weak. 

We now proceed to the proof of Theorem 1.
The translation-invariant distribution (or measure) $\nu\kappa_Jw_\Gamma$ is for triples $(J,\Gamma,\Psi)$; here we use 
$\Psi$ (as well as $\Gamma$) to denote an arbitrary state, and express $w_\Gamma$ as $w_\Gamma(\Psi)$, such 
that $\int_{\Psi\in A} dw_\Gamma(\Psi) =\int_{\Gamma_\alpha\in A}dw_\Gamma(\alpha)$ for any measurable set $A$. 
In the space of pairs $(J,\Psi)$ consisting of a bond realization and a state, we consider (Borel measurable) invariant sets 
$A$ of pairs; that is, if $(J,\Psi)\in A$, and $\theta$ is any translation or local transformation, then $(\theta J,\theta \Psi)
\in A$. These sets form a sub--$\sigma$-algebra ${\cal I}_1$ 
of the $\sigma$-algebra of all Borel sets of pairs. For a set $A\in{\cal I}_1$, we write $A_J$ for the set of $\Psi$ at the 
specified $J$; then $A_J$ changes covariantly under either a translation or a local change in $J$. (We will later
connect these sets with the invariant observables already discussed.) For the spin-flip invariant case, the definition of 
${\cal I}_1$ is modified because the local transformations are restricted to $|X|$ even, and further we impose the condition 
that for sets $A$ in ${\cal I}_1$, if $(J,\Psi)\in A$ then $(J,\overline{\Psi})\in A$.

The formal statement we prove is the following zero-one law, which is equivalent to the Theorem; after its proof we 
explain why that is so.

{\bf Proposition 1} (zero-one law): Consider a mixed $p$-spin model as in the hypotheses of Theorem 1, a metastate 
$\kappa_J$, and sets $A\in{\cal I}_1$. Then for $\nu(J)\kappa_J(\Gamma)$-almost every 
$(J,\Gamma)$, the measure $w_\Gamma$ is trivial on the sets $A_J$:  any such set has $w_\Gamma$-measure either 
$0$ or $1$.

{\bf Proof}: First consider the case without spin-flip symmetry. The $\kappa_J$-expectation 
of the measure $w_{\Gamma}(A_J)$ of the set $A_J$ for given $\Gamma$ is
\be
{\bf E}_{\kappa_J}
\int_{A_J} dw_{\Gamma}(\Psi).
\label{measAJ}
\ee 
By the translation covariance of $\kappa_J$, $w_\Gamma$, 
and $A_J$, this quantity is translation invariant. Hence as the distribution $\nu(J)$ is translation ergodic, (\ref{measAJ}) 
must be constant, i.e., independent of $J$ for $\nu$-almost every $J$ \cite{Georgii88}. 
On the other hand, for any $X$, under a local transformation in which only $J_X$ changes (by $\Delta J_X$), 
to first order (\ref{measAJ}) changes by
\be
\beta\Delta J_X {\bf E}_{\kappa_J}\int_{A_J} dw_\Gamma(\Psi)[\langle s_X\rangle_{\Psi}
-\langle s_X\rangle_\Gamma],
\label{DelJexp}
\ee
using~(\ref{wloctran}) and covariance of $\kappa_J$ and $A_J$. By the n.i.p.\ property, for $\nu$-almost every
given $J_X$ there is nonzero marginal probability for sets of $J_X'\neq J_X$ with $J_X'$ close to $J_X$, so from ergodicity, 
(\ref{DelJexp}) must be zero. 
Then applying the pure-state decomposition $\Gamma=
\int dw_\Gamma(\Psi)\Psi$, we have
\be
{\bf E}_{\kappa_J}\!\!\int_{A_J} \! dw_\Gamma(\Psi)\langle s_X\rangle_{\Psi}
={\bf E}_{\kappa_J}\!\!\int_{A_J} \!\! dw_\Gamma(\Psi)\!\int\! dw_\Gamma(\Psi')\langle s_X\rangle_{\Psi'}.
\label{zoid}
\ee
As~(\ref{zoid}) holds for all $X$, it follows that the states on the two sides are equal (relabeling $\Psi$ as $\Psi'$ 
on the left):
\be
{\bf E}_{\kappa_J}\!\int_{A_J} \! dw_\Gamma(\Psi')\Psi'
={\bf E}_{\kappa_J}\!\int_{A_J} \! dw_\Gamma(\Psi)\int dw_\Gamma(\Psi')\Psi'.
\ee
Both sides are weighted averages of pure states using a probability measure so, if nonzero, are Gibbs states up to normalization. 
Now the uniqueness of the pure-state decomposition of any Gibbs state for a given $J$ implies that 
there is a contradiction unless the measures on $\Psi'$ on the two sides are the same. In particular, 
as $A_J$ is independent of $\Gamma$, and on the left-hand side only a $\Psi'$ in $A_J$ can contribute, there is a 
contradiction unless on the right-hand side $\Psi'$ almost always lies in $A_J$. This implies that $\kappa_J$-almost every 
Gibbs state $\Gamma$ that has nonzero $w_\Gamma$-measure for $A_J$ must in fact have $w_\Gamma$-measure one
for $A_J$, $\int_{A_J} dw_\Gamma(\Psi)=1$; 
further this holds for $\nu$-almost every $J$. Finally, for the case with spin-flip symmetry, the proof is identical except that 
$|X|$ is even. $\Box$ 

Thus it is impossible to ``separate'' pure states into complementary covariant sets $A_J$, $A_J^c$ 
for $A\in{\cal I}_1$ that both have nonzero $w_\Gamma$ measure. Theorem 1 follows, because an invariant observable 
that took different values on (disjoint Borel) sets of pure states, each with nonzero $w_\Gamma$ measure, would 
thereby produce such a pair of sets. Conversely, for any set $A\in{\cal I}_1$, there is a (Borel measurable) function 
on pairs $(J,\Psi)$ that is equal to $1$ on $A$ and zero otherwise, and is invariant, 
so that the Theorem implies the Proposition. The observables in the examples
may not all be defined for all $\Psi$, but it is sufficient that all are well-defined for 
$\nu{\bf E}_{\kappa_J}w_\Gamma$-almost every pair $(J,\Psi)$. 
The proof of the result can be easily extended to give similar 
statements for more general spins and symmetries of their Hamiltonian. 

To conclude, in a broad class of models we have established a property, single-replica equivalence, 
of nontrivial mixed Gibbs states for short-range spin systems with disorder within any fully-covariant metastate construction;
it asserts that, for any Gibbs state drawn from the metastate for given disorder, each macroscopic observable takes 
the same value in any pure state in the decomposition of that Gibbs state. As discussed above, this considerably extends 
older results \cite{hb,AW90,chatterjee_15} that constrain the possible structure of mixed Gibbs states in such a metastate in a
disordered spin system, including at a first-order transition point. The result for self-overlaps was used 
as an assumption in rigorous proofs of other results \cite{NS09,ANS15}; it parallels a result \cite{Panchenko10a} for the SK 
model \cite{SK75}. The case of parts of the internal energy density leads directly to a set of identities of the stochastic-stability 
type \cite{AC98} via methods of Ref.\ \cite{GG98}. Such identities, as well as those of Ref.\ \cite{ANS15}, could play a key future
role in further constraining the behavior of such systems, similar to the case of the SK model \cite{panch}. These applications 
illustrate the significance of this fundamental general principle, proved here.


{\bf Acknowledgments}: We thank A.C.D. van Enter for thoughtful comments on an earlier version. 
NR is grateful for the support of NSF grant no.\ DMR-1724923.

\bibliography{refs.bib}

\small\def\em{\it} \newcommand{\noopsort}[1]{} \newcommand{\printfirst}[2]{#1}
  \newcommand{\singleletter}[1]{#1} \newcommand{\switchargs}[2]{#2#1}
\begin{thebibliography}{47}%
\makeatletter
\providecommand \@ifxundefined [1]{%
 \@ifx{#1\undefined}
}%
\providecommand \@ifnum [1]{%
 \ifnum #1\expandafter \@firstoftwo
 \else \expandafter \@secondoftwo
 \fi
}%
\providecommand \@ifx [1]{%
 \ifx #1\expandafter \@firstoftwo
 \else \expandafter \@secondoftwo
 \fi
}%
\providecommand \natexlab [1]{#1}%
\providecommand \enquote  [1]{``#1''}%
\providecommand \bibnamefont  [1]{#1}%
\providecommand \bibfnamefont [1]{#1}%
\providecommand \citenamefont [1]{#1}%
\providecommand \href@noop [0]{\@secondoftwo}%
\providecommand \href [0]{\begingroup \@sanitize@url \@href}%
\providecommand \@href[1]{\@@startlink{#1}\@@href}%
\providecommand \@@href[1]{\endgroup#1\@@endlink}%
\providecommand \@sanitize@url [0]{\catcode `\\12\catcode `\$12\catcode
  `\&12\catcode `\#12\catcode `\^12\catcode `\_12\catcode `\%12\relax}%
\providecommand \@@startlink[1]{}%
\providecommand \@@endlink[0]{}%
\providecommand \url  [0]{\begingroup\@sanitize@url \@url }%
\providecommand \@url [1]{\endgroup\@href {#1}{\urlprefix }}%
\providecommand \urlprefix  [0]{URL }%
\providecommand \Eprint [0]{\href }%
\providecommand \doibase [0]{http://dx.doi.org/}%
\providecommand \selectlanguage [0]{\@gobble}%
\providecommand \bibinfo  [0]{\@secondoftwo}%
\providecommand \bibfield  [0]{\@secondoftwo}%
\providecommand \translation [1]{[#1]}%
\providecommand \BibitemOpen [0]{}%
\providecommand \bibitemStop [0]{}%
\providecommand \bibitemNoStop [0]{.\EOS\space}%
\providecommand \EOS [0]{\spacefactor3000\relax}%
\providecommand \BibitemShut  [1]{\csname bibitem#1\endcsname}%
\let\auto@bib@innerbib\@empty
\bibitem [{\citenamefont {Aizenman}\ and\ \citenamefont
  {Contucci}(1998)}]{AC98}%
  \BibitemOpen
  \bibfield  {author} {\bibinfo {author} {\bibfnamefont {M.}~\bibnamefont
  {Aizenman}}\ and\ \bibinfo {author} {\bibfnamefont {P.}~\bibnamefont
  {Contucci}},\ }\href@noop {} {\bibfield  {journal} {\bibinfo  {journal} {J.
  Stat. Phys.}\ }\textbf {\bibinfo {volume} {92}},\ \bibinfo {pages} {765}
  (\bibinfo {year} {1998})}\BibitemShut {NoStop}%
\bibitem [{\citenamefont {Ghirlanda}\ and\ \citenamefont
  {Guerra}(1998)}]{GG98}%
  \BibitemOpen
  \bibfield  {author} {\bibinfo {author} {\bibfnamefont {S.}~\bibnamefont
  {Ghirlanda}}\ and\ \bibinfo {author} {\bibfnamefont {F.}~\bibnamefont
  {Guerra}},\ }\href@noop {} {\bibfield  {journal} {\bibinfo  {journal} {J.
  Phys. A}\ }\textbf {\bibinfo {volume} {31}},\ \bibinfo {pages} {9149}
  (\bibinfo {year} {1998})}\BibitemShut {NoStop}%
\bibitem [{\citenamefont {Guerra}\ and\ \citenamefont
  {Toninelli}(2002)}]{GT02}%
  \BibitemOpen
  \bibfield  {author} {\bibinfo {author} {\bibfnamefont {F.}~\bibnamefont
  {Guerra}}\ and\ \bibinfo {author} {\bibfnamefont {F.~L.}\ \bibnamefont
  {Toninelli}},\ }\href@noop {} {\bibfield  {journal} {\bibinfo  {journal}
  {Commun. Math. Phys.}\ }\textbf {\bibinfo {volume} {230}},\ \bibinfo {pages}
  {71} (\bibinfo {year} {2002})}\BibitemShut {NoStop}%
\bibitem [{\citenamefont {Guerra}(2003)}]{Guerra03}%
  \BibitemOpen
  \bibfield  {author} {\bibinfo {author} {\bibfnamefont {F.}~\bibnamefont
  {Guerra}},\ }\href@noop {} {\bibfield  {journal} {\bibinfo  {journal}
  {Commun. Math. Phys.}\ }\textbf {\bibinfo {volume} {233}},\ \bibinfo {pages}
  {1} (\bibinfo {year} {2003})}\BibitemShut {NoStop}%
\bibitem [{\citenamefont {Talagrand}(rlin)}]{Talagrand03a}%
  \BibitemOpen
  \bibfield  {author} {\bibinfo {author} {\bibfnamefont {M.}~\bibnamefont
  {Talagrand}},\ }\href@noop {} {\emph {\bibinfo {title} {Spin Glasses: A
  Challenge for Mathematicians}}}\ (\bibinfo  {publisher} {Springer-Verlag},\
  \bibinfo {address} {2003},\ \bibinfo {year} {Berlin})\BibitemShut {NoStop}%
\bibitem [{\citenamefont {Talagrand}(2003)}]{Talagrand03b}%
  \BibitemOpen
  \bibfield  {author} {\bibinfo {author} {\bibfnamefont {M.}~\bibnamefont
  {Talagrand}},\ }\href@noop {} {\bibfield  {journal} {\bibinfo  {journal} {C.
  R. Acad. Sci. Paris}\ }\textbf {\bibinfo {volume} {337}},\ \bibinfo {pages}
  {111} (\bibinfo {year} {2003})}\BibitemShut {NoStop}%
\bibitem [{\citenamefont {Talagrand}(2005)}]{Talagrand05}%
  \BibitemOpen
  \bibfield  {author} {\bibinfo {author} {\bibfnamefont {M.}~\bibnamefont
  {Talagrand}},\ }\href@noop {} {\bibfield  {journal} {\bibinfo  {journal}
  {Ann. Math.}\ }\textbf {\bibinfo {volume} {163}},\ \bibinfo {pages} {221}
  (\bibinfo {year} {2005})}\BibitemShut {NoStop}%
\bibitem [{\citenamefont {Panchenko}(2013)}]{panch}%
  \BibitemOpen
  \bibfield  {author} {\bibinfo {author} {\bibfnamefont {D.}~\bibnamefont
  {Panchenko}},\ }\href@noop {} {\bibfield  {journal} {\bibinfo  {journal}
  {Ann. Math.}\ }\textbf {\bibinfo {volume} {177}},\ \bibinfo {pages} {383}
  (\bibinfo {year} {2013})}\BibitemShut {NoStop}%
\bibitem [{\citenamefont {Newman}\ \emph {et~al.}(2022)\citenamefont {Newman},
  \citenamefont {Read},\ and\ \citenamefont {Stein}}]{NRS22}%
  \BibitemOpen
  \bibfield  {author} {\bibinfo {author} {\bibfnamefont {C.~M.}\ \bibnamefont
  {Newman}}, \bibinfo {author} {\bibfnamefont {N.}~\bibnamefont {Read}}, \ and\
  \bibinfo {author} {\bibfnamefont {D.~L.}\ \bibnamefont {Stein}},\ }in\
  \href@noop {} {\emph {\bibinfo {booktitle} {Spin Glass Theory and Far Beyond
  --- Replica Symmetry Breaking after 40 Years}}},\ \bibinfo {editor} {edited
  by\ \bibinfo {editor} {\bibfnamefont {P.}~\bibnamefont {Charbonneau}},
  \bibinfo {editor} {\bibfnamefont {E.}~\bibnamefont {Marinari}}, \bibinfo
  {editor} {\bibfnamefont {M.}~\bibnamefont {M\'ezard}}, \bibinfo {editor}
  {\bibfnamefont {F.}~\bibnamefont {Ricci-Tersenghi}}, \bibinfo {editor}
  {\bibfnamefont {G.}~\bibnamefont {Sicuro}}, \ and\ \bibinfo {editor}
  {\bibfnamefont {F.}~\bibnamefont {Zamponi}}}\ (\bibinfo  {publisher} {World
  Scientific},\ \bibinfo {year} {2022})\BibitemShut {NoStop}%
\bibitem [{\citenamefont {Sherrington}\ and\ \citenamefont
  {Kirkpatrick}(1975)}]{SK75}%
  \BibitemOpen
  \bibfield  {author} {\bibinfo {author} {\bibfnamefont {D.}~\bibnamefont
  {Sherrington}}\ and\ \bibinfo {author} {\bibfnamefont {S.}~\bibnamefont
  {Kirkpatrick}},\ }\href@noop {} {\bibfield  {journal} {\bibinfo  {journal}
  {Phys. Rev. Lett.}\ }\textbf {\bibinfo {volume} {35}},\ \bibinfo {pages}
  {1792} (\bibinfo {year} {1975})}\BibitemShut {NoStop}%
\bibitem [{\citenamefont {Parisi}(1979)}]{Parisi79}%
  \BibitemOpen
  \bibfield  {author} {\bibinfo {author} {\bibfnamefont {G.}~\bibnamefont
  {Parisi}},\ }\href@noop {} {\bibfield  {journal} {\bibinfo  {journal} {Phys.
  Rev. Lett.}\ }\textbf {\bibinfo {volume} {43}},\ \bibinfo {pages} {1754}
  (\bibinfo {year} {1979})}\BibitemShut {NoStop}%
\bibitem [{\citenamefont {Parisi}(1983)}]{Parisi83}%
  \BibitemOpen
  \bibfield  {author} {\bibinfo {author} {\bibfnamefont {G.}~\bibnamefont
  {Parisi}},\ }\href@noop {} {\bibfield  {journal} {\bibinfo  {journal} {Phys.
  Rev. Lett.}\ }\textbf {\bibinfo {volume} {50}},\ \bibinfo {pages} {1946}
  (\bibinfo {year} {1983})}\BibitemShut {NoStop}%
\bibitem [{\citenamefont {M\'ezard}\ \emph
  {et~al.}(1984{\natexlab{a}})\citenamefont {M\'ezard}, \citenamefont {Parisi},
  \citenamefont {Sourlas}, \citenamefont {Toulouse},\ and\ \citenamefont
  {Virasoro}}]{MPSTV84a}%
  \BibitemOpen
  \bibfield  {author} {\bibinfo {author} {\bibfnamefont {M.}~\bibnamefont
  {M\'ezard}}, \bibinfo {author} {\bibfnamefont {G.}~\bibnamefont {Parisi}},
  \bibinfo {author} {\bibfnamefont {N.}~\bibnamefont {Sourlas}}, \bibinfo
  {author} {\bibfnamefont {G.}~\bibnamefont {Toulouse}}, \ and\ \bibinfo
  {author} {\bibfnamefont {M.}~\bibnamefont {Virasoro}},\ }\href@noop {}
  {\bibfield  {journal} {\bibinfo  {journal} {Phys. Rev. Lett.}\ }\textbf
  {\bibinfo {volume} {52}},\ \bibinfo {pages} {1156} (\bibinfo {year}
  {1984}{\natexlab{a}})}\BibitemShut {NoStop}%
\bibitem [{\citenamefont {M\'ezard}\ \emph
  {et~al.}(1984{\natexlab{b}})\citenamefont {M\'ezard}, \citenamefont {Parisi},
  \citenamefont {Sourlas}, \citenamefont {Toulouse},\ and\ \citenamefont
  {Virasoro}}]{MPSTV84b}%
  \BibitemOpen
  \bibfield  {author} {\bibinfo {author} {\bibfnamefont {M.}~\bibnamefont
  {M\'ezard}}, \bibinfo {author} {\bibfnamefont {G.}~\bibnamefont {Parisi}},
  \bibinfo {author} {\bibfnamefont {N.}~\bibnamefont {Sourlas}}, \bibinfo
  {author} {\bibfnamefont {G.}~\bibnamefont {Toulouse}}, \ and\ \bibinfo
  {author} {\bibfnamefont {M.}~\bibnamefont {Virasoro}},\ }\href@noop {}
  {\bibfield  {journal} {\bibinfo  {journal} {J. Phys. (Paris)}\ }\textbf
  {\bibinfo {volume} {45}},\ \bibinfo {pages} {843} (\bibinfo {year}
  {1984}{\natexlab{b}})}\BibitemShut {NoStop}%
\bibitem [{\citenamefont {M\'ezard}\ \emph {et~al.}(1987)\citenamefont
  {M\'ezard}, \citenamefont {Parisi},\ and\ \citenamefont {Virasoro}}]{MPV87}%
  \BibitemOpen
  \bibinfo {editor} {\bibfnamefont {M.}~\bibnamefont {M\'ezard}}, \bibinfo
  {editor} {\bibfnamefont {G.}~\bibnamefont {Parisi}}, \ and\ \bibinfo {editor}
  {\bibfnamefont {M.~A.}\ \bibnamefont {Virasoro}},\ eds.,\ \href@noop {}
  {\emph {\bibinfo {title} {Spin Glass Theory and Beyond}}}\ (\bibinfo
  {publisher} {World Scientific},\ \bibinfo {address} {Singapore},\ \bibinfo
  {year} {1987})\BibitemShut {NoStop}%
\bibitem [{\citenamefont {McMillan}(1984)}]{Mac84}%
  \BibitemOpen
  \bibfield  {author} {\bibinfo {author} {\bibfnamefont {W.~L.}\ \bibnamefont
  {McMillan}},\ }\href@noop {} {\bibfield  {journal} {\bibinfo  {journal} {J.
  Phys. C}\ }\textbf {\bibinfo {volume} {17}},\ \bibinfo {pages} {3179}
  (\bibinfo {year} {1984})}\BibitemShut {NoStop}%
\bibitem [{\citenamefont {Bray}\ and\ \citenamefont {Moore}(1985)}]{BM85}%
  \BibitemOpen
  \bibfield  {author} {\bibinfo {author} {\bibfnamefont {A.~J.}\ \bibnamefont
  {Bray}}\ and\ \bibinfo {author} {\bibfnamefont {M.~A.}\ \bibnamefont
  {Moore}},\ }\href@noop {} {\bibfield  {journal} {\bibinfo  {journal} {Phys.
  Rev. B}\ }\textbf {\bibinfo {volume} {31}},\ \bibinfo {pages} {631} (\bibinfo
  {year} {1985})}\BibitemShut {NoStop}%
\bibitem [{\citenamefont {Fisher}\ and\ \citenamefont {Huse}(1988)}]{FH88b}%
  \BibitemOpen
  \bibfield  {author} {\bibinfo {author} {\bibfnamefont {D.~S.}\ \bibnamefont
  {Fisher}}\ and\ \bibinfo {author} {\bibfnamefont {D.~A.}\ \bibnamefont
  {Huse}},\ }\href@noop {} {\bibfield  {journal} {\bibinfo  {journal} {Phys.
  Rev. B}\ }\textbf {\bibinfo {volume} {38}},\ \bibinfo {pages} {386} (\bibinfo
  {year} {1988})}\BibitemShut {NoStop}%
\bibitem [{\citenamefont {Fisher}\ and\ \citenamefont {Huse}(1986)}]{FH86}%
  \BibitemOpen
  \bibfield  {author} {\bibinfo {author} {\bibfnamefont {D.~S.}\ \bibnamefont
  {Fisher}}\ and\ \bibinfo {author} {\bibfnamefont {D.~A.}\ \bibnamefont
  {Huse}},\ }\href@noop {} {\bibfield  {journal} {\bibinfo  {journal} {Phys.
  Rev. Lett.}\ }\textbf {\bibinfo {volume} {56}},\ \bibinfo {pages} {1601}
  (\bibinfo {year} {1986})}\BibitemShut {NoStop}%
\bibitem [{\citenamefont {Aizenman}\ and\ \citenamefont {Wehr}(1990)}]{AW90}%
  \BibitemOpen
  \bibfield  {author} {\bibinfo {author} {\bibfnamefont {M.}~\bibnamefont
  {Aizenman}}\ and\ \bibinfo {author} {\bibfnamefont {J.}~\bibnamefont
  {Wehr}},\ }\href@noop {} {\bibfield  {journal} {\bibinfo  {journal} {Commun.
  Math. Phys.}\ }\textbf {\bibinfo {volume} {130}},\ \bibinfo {pages} {489}
  (\bibinfo {year} {1990})}\BibitemShut {NoStop}%
\bibitem [{\citenamefont {Newman}\ and\ \citenamefont {Stein}(1996)}]{NS96c}%
  \BibitemOpen
  \bibfield  {author} {\bibinfo {author} {\bibfnamefont {C.~M.}\ \bibnamefont
  {Newman}}\ and\ \bibinfo {author} {\bibfnamefont {D.~L.}\ \bibnamefont
  {Stein}},\ }\href@noop {} {\bibfield  {journal} {\bibinfo  {journal} {Phys.
  Rev. Lett.}\ }\textbf {\bibinfo {volume} {76}},\ \bibinfo {pages} {4821}
  (\bibinfo {year} {1996})}\BibitemShut {NoStop}%
\bibitem [{\citenamefont {Newman}\ and\ \citenamefont {Stein}(1997)}]{NS97}%
  \BibitemOpen
  \bibfield  {author} {\bibinfo {author} {\bibfnamefont {C.~M.}\ \bibnamefont
  {Newman}}\ and\ \bibinfo {author} {\bibfnamefont {D.~L.}\ \bibnamefont
  {Stein}},\ }\href@noop {} {\bibfield  {journal} {\bibinfo  {journal} {Phys.
  Rev. E}\ }\textbf {\bibinfo {volume} {55}},\ \bibinfo {pages} {5194}
  (\bibinfo {year} {1997})}\BibitemShut {NoStop}%
\bibitem [{\citenamefont {Newman}\ and\ \citenamefont
  {Stein}(1998{\natexlab{a}})}]{NSBerlin}%
  \BibitemOpen
  \bibfield  {author} {\bibinfo {author} {\bibfnamefont {C.~M.}\ \bibnamefont
  {Newman}}\ and\ \bibinfo {author} {\bibfnamefont {D.~L.}\ \bibnamefont
  {Stein}},\ }in\ \href@noop {} {\emph {\bibinfo {booktitle} {Mathematics of
  Spin Glasses and Neural Networks}}},\ \bibinfo {editor} {edited by\ \bibinfo
  {editor} {\bibfnamefont {A.}~\bibnamefont {Bovier}}\ and\ \bibinfo {editor}
  {\bibfnamefont {P.}~\bibnamefont {Picco}}}\ (\bibinfo  {publisher}
  {Birkhauser},\ \bibinfo {address} {Boston},\ \bibinfo {year} {1998})\ pp.\
  \bibinfo {pages} {243--287}\BibitemShut {NoStop}%
\bibitem [{\citenamefont {Newman}\ and\ \citenamefont {Stein}(2002)}]{NS02}%
  \BibitemOpen
  \bibfield  {author} {\bibinfo {author} {\bibfnamefont {C.~M.}\ \bibnamefont
  {Newman}}\ and\ \bibinfo {author} {\bibfnamefont {D.~L.}\ \bibnamefont
  {Stein}},\ }\href@noop {} {\bibfield  {journal} {\bibinfo  {journal} {J.
  Stat. Phys.}\ }\textbf {\bibinfo {volume} {106}},\ \bibinfo {pages} {213}
  (\bibinfo {year} {2002})}\BibitemShut {NoStop}%
\bibitem [{\citenamefont {Newman}\ and\ \citenamefont {Stein}(2003)}]{NS03b}%
  \BibitemOpen
  \bibfield  {author} {\bibinfo {author} {\bibfnamefont {C.~M.}\ \bibnamefont
  {Newman}}\ and\ \bibinfo {author} {\bibfnamefont {D.~L.}\ \bibnamefont
  {Stein}},\ }\href@noop {} {\bibfield  {journal} {\bibinfo  {journal} {J.
  Phys.: Cond. Mat.}\ }\textbf {\bibinfo {volume} {15}},\ \bibinfo {pages}
  {R1319 } (\bibinfo {year} {2003})}\BibitemShut {NoStop}%
\bibitem [{\citenamefont {Read}(2014)}]{Read14}%
  \BibitemOpen
  \bibfield  {author} {\bibinfo {author} {\bibfnamefont {N.}~\bibnamefont
  {Read}},\ }\href@noop {} {\bibfield  {journal} {\bibinfo  {journal} {Phys.
  Rev. E}\ }\textbf {\bibinfo {volume} {90}},\ \bibinfo {pages} {032142}
  (\bibinfo {year} {2014})}\BibitemShut {NoStop}%
\bibitem [{\citenamefont {Panchenko}(2010)}]{Panchenko10a}%
  \BibitemOpen
  \bibfield  {author} {\bibinfo {author} {\bibfnamefont {D.}~\bibnamefont
  {Panchenko}},\ }\href@noop {} {\bibfield  {journal} {\bibinfo  {journal}
  {Ann. Prob.}\ }\textbf {\bibinfo {volume} {38}},\ \bibinfo {pages} {327}
  (\bibinfo {year} {2010})}\BibitemShut {NoStop}%
\bibitem [{\citenamefont {Parisi}(2004)}]{Parisi04}%
  \BibitemOpen
  \bibfield  {author} {\bibinfo {author} {\bibfnamefont {G.}~\bibnamefont
  {Parisi}},\ }\href@noop {} {\bibfield  {journal} {\bibinfo  {journal} {Int.
  J. Modern Phys. A}\ }\textbf {\bibinfo {volume} {18}},\ \bibinfo {pages}
  {733} (\bibinfo {year} {2004})}\BibitemShut {NoStop}%
\bibitem [{\citenamefont {Read}(2022)}]{Read22}%
  \BibitemOpen
  \bibfield  {author} {\bibinfo {author} {\bibfnamefont {N.}~\bibnamefont
  {Read}},\ }\href@noop {} {\bibfield  {journal} {\bibinfo  {journal} {Phys.
  Rev. E}\ }\textbf {\bibinfo {volume} {105}},\ \bibinfo {pages} {054134}
  (\bibinfo {year} {2022})}\BibitemShut {NoStop}%
\bibitem [{\citenamefont {Edwards}\ and\ \citenamefont
  {Anderson}(1975)}]{EA75}%
  \BibitemOpen
  \bibfield  {author} {\bibinfo {author} {\bibfnamefont {S.}~\bibnamefont
  {Edwards}}\ and\ \bibinfo {author} {\bibfnamefont {P.~W.}\ \bibnamefont
  {Anderson}},\ }\href@noop {} {\bibfield  {journal} {\bibinfo  {journal} {J.
  Phys. F}\ }\textbf {\bibinfo {volume} {5}},\ \bibinfo {pages} {965} (\bibinfo
  {year} {1975})}\BibitemShut {NoStop}%
\bibitem [{\citenamefont {Georgii}(1988)}]{Georgii88}%
  \BibitemOpen
  \bibfield  {author} {\bibinfo {author} {\bibfnamefont {H.~O.}\ \bibnamefont
  {Georgii}},\ }\href@noop {} {\emph {\bibinfo {title} {Gibbs Measures and
  Phase Transitions}}}\ (\bibinfo  {publisher} {de Gruyter},\ \bibinfo
  {address} {Berlin},\ \bibinfo {year} {1988})\BibitemShut {NoStop}%
\bibitem [{\citenamefont {Simon}(1993)}]{Simon92}%
  \BibitemOpen
  \bibfield  {author} {\bibinfo {author} {\bibfnamefont {B.}~\bibnamefont
  {Simon}},\ }\href@noop {} {\emph {\bibinfo {title} {The Statistical Mechanics
  of Lattice Gases}}}\ (\bibinfo  {publisher} {Princeton University Press},\
  \bibinfo {address} {Princeton, NJ},\ \bibinfo {year} {1993})\BibitemShut
  {NoStop}%
\bibitem [{\citenamefont {Newman}\ and\ \citenamefont
  {Stein}(1998{\natexlab{b}})}]{NS98}%
  \BibitemOpen
  \bibfield  {author} {\bibinfo {author} {\bibfnamefont {C.~M.}\ \bibnamefont
  {Newman}}\ and\ \bibinfo {author} {\bibfnamefont {D.~L.}\ \bibnamefont
  {Stein}},\ }\href@noop {} {\bibfield  {journal} {\bibinfo  {journal} {Phys.
  Rev. E}\ }\textbf {\bibinfo {volume} {57}},\ \bibinfo {pages} {1356}
  (\bibinfo {year} {1998}{\natexlab{b}})}\BibitemShut {NoStop}%
\bibitem [{\citenamefont {Newman}\ and\ \citenamefont {Stein}(1992)}]{NS92}%
  \BibitemOpen
  \bibfield  {author} {\bibinfo {author} {\bibfnamefont {C.~M.}\ \bibnamefont
  {Newman}}\ and\ \bibinfo {author} {\bibfnamefont {D.~L.}\ \bibnamefont
  {Stein}},\ }\href@noop {} {\bibfield  {journal} {\bibinfo  {journal} {Phys.
  Rev. B}\ }\textbf {\bibinfo {volume} {46}},\ \bibinfo {pages} {973} (\bibinfo
  {year} {1992})}\BibitemShut {NoStop}%
\bibitem [{\citenamefont {Imry}\ and\ \citenamefont {Ma}(1975)}]{IM75}%
  \BibitemOpen
  \bibfield  {author} {\bibinfo {author} {\bibfnamefont {Y.}~\bibnamefont
  {Imry}}\ and\ \bibinfo {author} {\bibfnamefont {S.-K.}\ \bibnamefont {Ma}},\
  }\href@noop {} {\bibfield  {journal} {\bibinfo  {journal} {Phys. Rev. Lett.}\
  }\textbf {\bibinfo {volume} {35}},\ \bibinfo {pages} {1399} (\bibinfo {year}
  {1975})}\BibitemShut {NoStop}%
\bibitem [{\citenamefont {Chatterjee}(2015)}]{chatterjee_15}%
  \BibitemOpen
  \bibfield  {author} {\bibinfo {author} {\bibfnamefont {S.}~\bibnamefont
  {Chatterjee}},\ }\href@noop {} {\bibfield  {journal} {\bibinfo  {journal}
  {Commun. Math. Phys.}\ }\textbf {\bibinfo {volume} {337}},\ \bibinfo {pages}
  {93} (\bibinfo {year} {2015})}\BibitemShut {NoStop}%
\bibitem [{\citenamefont {Hui}\ and\ \citenamefont {Berker}(1989)}]{hb}%
  \BibitemOpen
  \bibfield  {author} {\bibinfo {author} {\bibfnamefont {K.}~\bibnamefont
  {Hui}}\ and\ \bibinfo {author} {\bibfnamefont {A.~N.}\ \bibnamefont
  {Berker}},\ }\href@noop {} {\bibfield  {journal} {\bibinfo  {journal} {Phys.
  Rev. Lett.}\ }\textbf {\bibinfo {volume} {62}},\ \bibinfo {pages} {2507}
  (\bibinfo {year} {1989})}\BibitemShut {NoStop}%
\bibitem [{\citenamefont {Harris}(1974)}]{harris}%
  \BibitemOpen
  \bibfield  {author} {\bibinfo {author} {\bibfnamefont {A.~B.}\ \bibnamefont
  {Harris}},\ }\href@noop {} {\bibfield  {journal} {\bibinfo  {journal} {J.
  Phys. C}\ }\textbf {\bibinfo {volume} {7}},\ \bibinfo {pages} {1671}
  (\bibinfo {year} {1974})}\BibitemShut {NoStop}%
\bibitem [{\citenamefont {Vuillermot}(1977)}]{vuillermot}%
  \BibitemOpen
  \bibfield  {author} {\bibinfo {author} {\bibfnamefont {P.~A.}\ \bibnamefont
  {Vuillermot}},\ }\href@noop {} {\bibfield  {journal} {\bibinfo  {journal} {J.
  Phys. A: Math. Gen.}\ }\textbf {\bibinfo {volume} {10}},\ \bibinfo {pages}
  {1319} (\bibinfo {year} {1977})}\BibitemShut {NoStop}%
\bibitem [{\citenamefont {Ledrappier}(1977)}]{ledrappier}%
  \BibitemOpen
  \bibfield  {author} {\bibinfo {author} {\bibfnamefont {F.}~\bibnamefont
  {Ledrappier}},\ }\href@noop {} {\bibfield  {journal} {\bibinfo  {journal}
  {Commun. Math. Phys.}\ }\textbf {\bibinfo {volume} {56}},\ \bibinfo {pages}
  {297} (\bibinfo {year} {1977})}\BibitemShut {NoStop}%
\bibitem [{\citenamefont {Pastur}\ and\ \citenamefont
  {Figotin}(1978)}]{pasfig}%
  \BibitemOpen
  \bibfield  {author} {\bibinfo {author} {\bibfnamefont {L.~A.}\ \bibnamefont
  {Pastur}}\ and\ \bibinfo {author} {\bibfnamefont {A.~L.}\ \bibnamefont
  {Figotin}},\ }\href@noop {} {\bibfield  {journal} {\bibinfo  {journal}
  {Theor. Math. Phys.}\ }\textbf {\bibinfo {volume} {35}},\ \bibinfo {pages}
  {403} (\bibinfo {year} {1978})}\BibitemShut {NoStop}%
\bibitem [{\citenamefont {van Hemmen}\ and\ \citenamefont
  {Palmer}(1982)}]{vHP}%
  \BibitemOpen
  \bibfield  {author} {\bibinfo {author} {\bibfnamefont {J.~L.}\ \bibnamefont
  {van Hemmen}}\ and\ \bibinfo {author} {\bibfnamefont {R.~G.}\ \bibnamefont
  {Palmer}},\ }\href@noop {} {\bibfield  {journal} {\bibinfo  {journal} {J.
  Phys. A: Math. Gen.}\ }\textbf {\bibinfo {volume} {15}},\ \bibinfo {pages}
  {3881} (\bibinfo {year} {1982})}\BibitemShut {NoStop}%
\bibitem [{\citenamefont {{A. C. D. van Enter and J. L. van
  Hemmen}}(1983)}]{vEvH_83}%
  \BibitemOpen
  \bibfield  {author} {\bibinfo {author} {\bibnamefont {{A. C. D. van Enter and
  J. L. van Hemmen}}},\ }\href@noop {} {\bibfield  {journal} {\bibinfo
  {journal} {J. Stat. Phys.}\ }\textbf {\bibinfo {volume} {32}},\ \bibinfo
  {pages} {141} (\bibinfo {year} {1983})}\BibitemShut {NoStop}%
\bibitem [{\citenamefont {Zegarlinski}(1991)}]{Zegarlinski91}%
  \BibitemOpen
  \bibfield  {author} {\bibinfo {author} {\bibfnamefont {B.}~\bibnamefont
  {Zegarlinski}},\ }\href@noop {} {\bibfield  {journal} {\bibinfo  {journal}
  {Commun. Math. Phys.}\ }\textbf {\bibinfo {volume} {139}},\ \bibinfo {pages}
  {305} (\bibinfo {year} {1991})}\BibitemShut {NoStop}%
\bibitem [{\citenamefont {Breiman}(1992)}]{breiman_book}%
  \BibitemOpen
  \bibfield  {author} {\bibinfo {author} {\bibfnamefont {L.}~\bibnamefont
  {Breiman}},\ }\href@noop {} {\emph {\bibinfo {title} {Probability}}}\
  (\bibinfo  {publisher} {Society for Industrial and Applied Mathematics},\
  \bibinfo {address} {Philadelphia},\ \bibinfo {year} {1992})\BibitemShut
  {NoStop}%
\bibitem [{\citenamefont {Newman}\ and\ \citenamefont {Stein}(2009)}]{NS09}%
  \BibitemOpen
  \bibfield  {author} {\bibinfo {author} {\bibfnamefont {C.~M.}\ \bibnamefont
  {Newman}}\ and\ \bibinfo {author} {\bibfnamefont {D.~L.}\ \bibnamefont
  {Stein}},\ }in\ \href@noop {} {\emph {\bibinfo {booktitle} {New Trends in
  Mathematical Physics: {Proceedings} of the 2006 International Congress of
  Mathematical Physics}}},\ \bibinfo {editor} {edited by\ \bibinfo {editor}
  {\bibfnamefont {V.}~\bibnamefont {Sidoravicius}}}\ (\bibinfo  {publisher}
  {Springer},\ \bibinfo {address} {New York},\ \bibinfo {year} {2009})\ pp.\
  \bibinfo {pages} {643--652}\BibitemShut {NoStop}%
\bibitem [{\citenamefont {Arguin}\ \emph {et~al.}(2015)\citenamefont {Arguin},
  \citenamefont {Newman},\ and\ \citenamefont {Stein}}]{ANS15}%
  \BibitemOpen
  \bibfield  {author} {\bibinfo {author} {\bibfnamefont {L.-P.}\ \bibnamefont
  {Arguin}}, \bibinfo {author} {\bibfnamefont {C.~M.}\ \bibnamefont {Newman}},
  \ and\ \bibinfo {author} {\bibfnamefont {D.~L.}\ \bibnamefont {Stein}},\
  }\href@noop {} {\bibfield  {journal} {\bibinfo  {journal} {Phys. Rev. Lett.}\
  }\textbf {\bibinfo {volume} {115}},\ \bibinfo {pages} {187202} (\bibinfo
  {year} {2015})}\BibitemShut {NoStop}%
\end{thebibliography}%

\end{document}